\documentclass[final,1p,times]{elsarticle} % do not change
\usepackage{graphicx} % do not change
\usepackage{amssymb} % do not change
\usepackage{amsthm} % do not change
\usepackage{lineno} % do not change

\newcommand{\gevc}{\mbox{GeV/$c$}~}

\newcommand{\pt}{\mbox{$p_{T}$}~}
\newcommand{\jpsi}{\mbox{$J/\psi~$}}

\newcommand{\psip}{\mbox{$\psi'$~}}

\newcommand{\ee}{\mbox{$e^{+}e^{-}$~}}

\newcommand{\pp}{\mbox{$p$+$p$~}}

\journal{Nuclear Physics A} % do not change
\begin{document} % do not change

\begin{frontmatter} % do not change

%% QM09Author: please enter your  
%% Title, author and address info here; please do not use footnotes

% Your Title - please insert
\title{Quarkonia measurement in $p$+$p$ and $d$+Au collisions at
  $\sqrt{s}$=200 GeV by PHENIX Detector.}

% Principle author, and co-authors - please insert
\author{Cesar Luiz da Silva for the PHENIX Collaboration}

% Address - please insert
\address{Iowa State University, Department of Physics and Astronomy,
  Ames, Iowa, 50011, USA}

\begin{abstract} % do not change
%% Text of abstract goes here - please insert
We report new quarkonia measurements necessary to understand
production mechanisms and cold nuclear matter effects in the yields
observed at RHIC energy. Results obtained in \pp collisions collected during
the 2006 RHIC Run include \jpsi, \psip and $\Upsilon$
differential cross sections as well as \jpsi polarization. Revisited
interpretations of the published \jpsi nuclear modification factors and
statistically improved observations in $d$+Au collisions taken in the 2008
Run are also discussed in the view of the recent understanding of the
initial state effects and breakup cross section.
\end{abstract} % do not change

\end{frontmatter} % do not change

%% QM09: we keep linenumbers at least for initial version
%\linenumbers % do not change

%% start of main text - please insert. 

\section{\pp baseline}
\label{sec:pp}

\begin{figure}[ht]
  \begin{minipage}{0.56\textwidth}
    % \centering
    \includegraphics[width=1.0\textwidth]{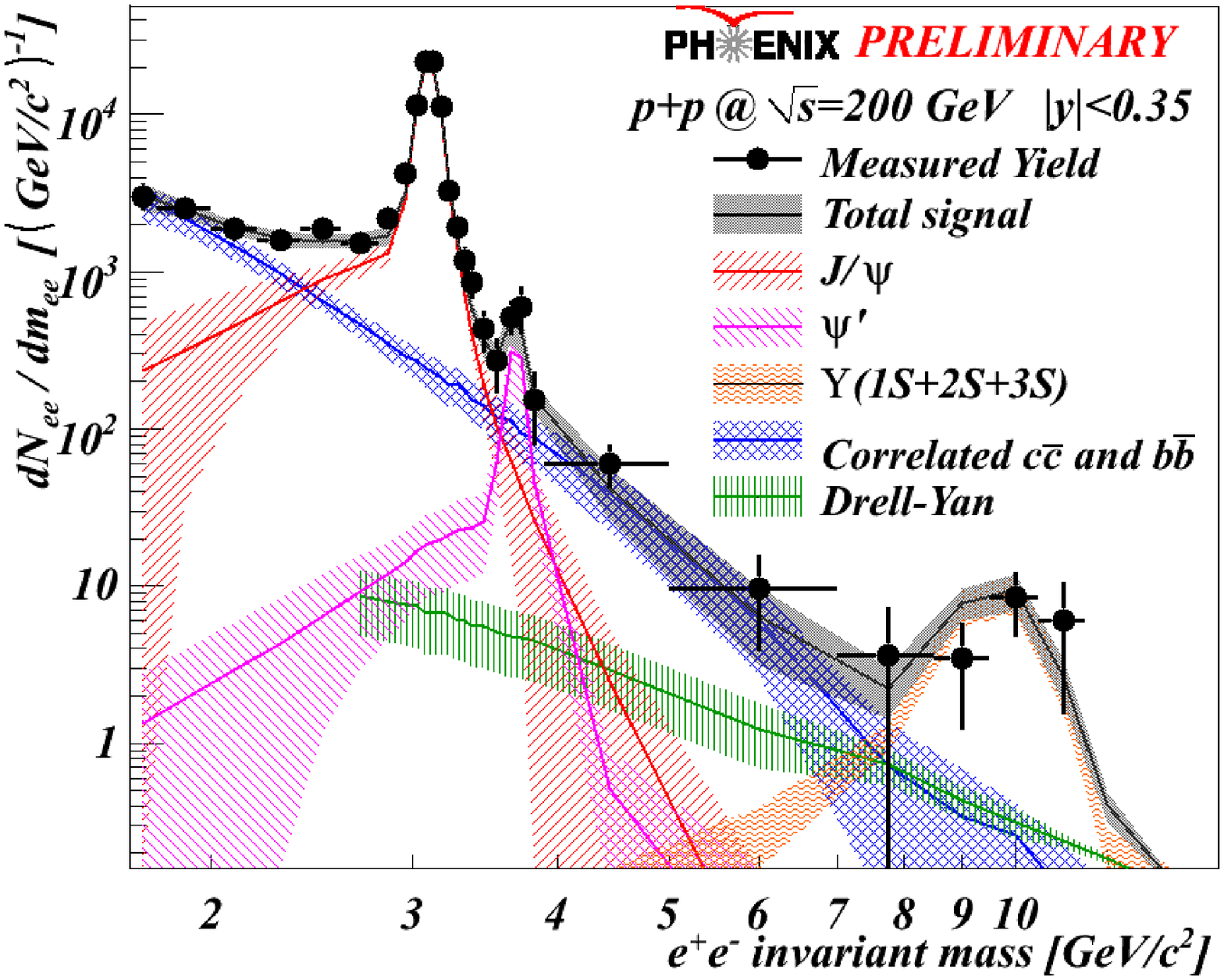}
    \caption{Combinatorial background subtracted invariant mass
      distribution of \ee and corresponding physical sources.}
    \label{invmass_fit}
  \end{minipage}\hfill
  \begin{minipage}{0.4\textwidth}
    % \centering
    \includegraphics[width=1.0\textwidth]{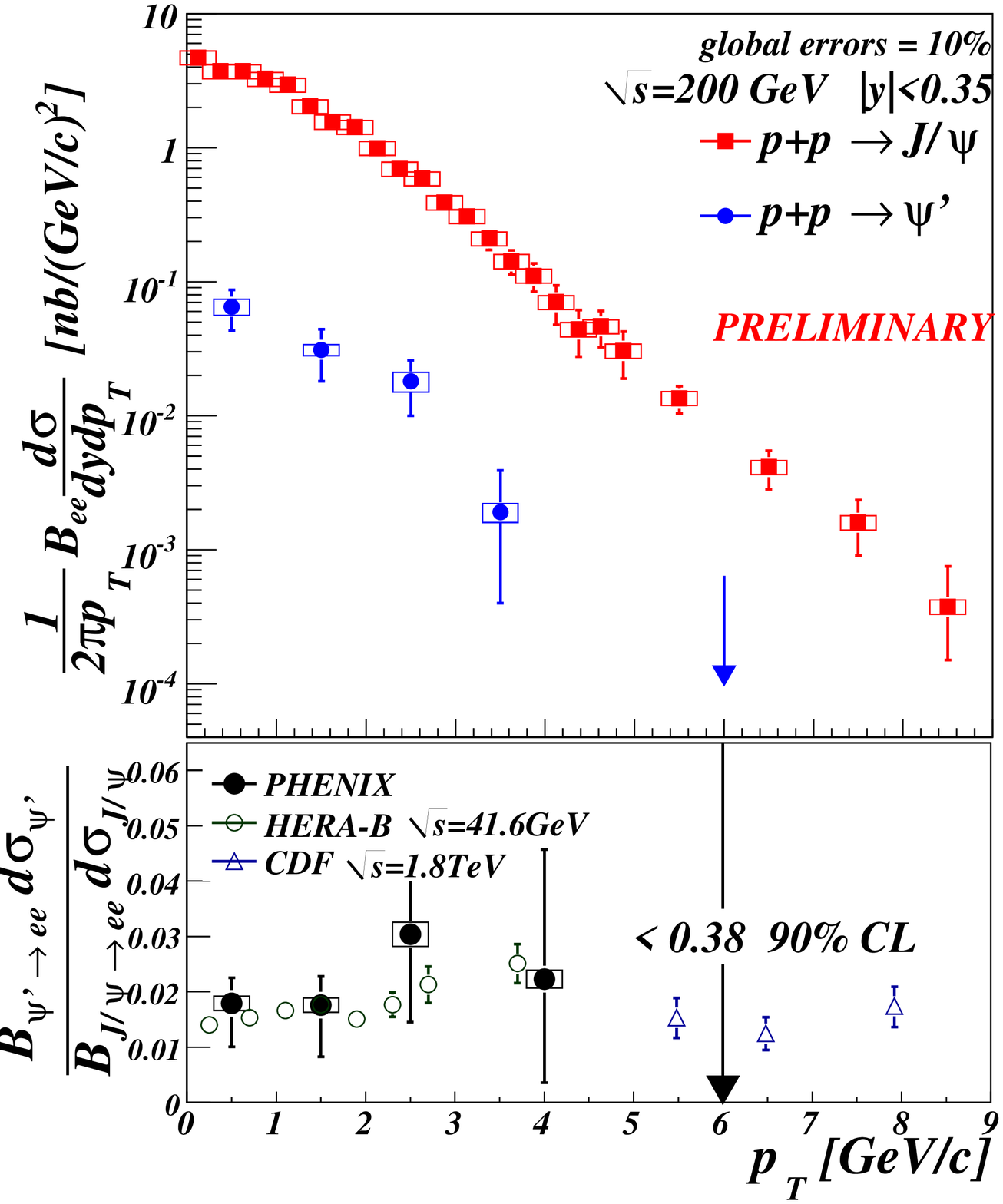}
    \caption{Transverse momentum dependence of \jpsi and \psip cross
      sections at mid-rapidity and corresponding \psip to \jpsi ratio
      together with that measured in HERA-B \cite{HERA-B} and CDF
      \cite{CDF}.}
    \label{jpsi_psip}
  \end{minipage}
\end{figure}

The invariant mass spectrum of dielectrons measured during the 2006
RHIC Run, after combinatorial background
subtraction (Figure \ref{invmass_fit}) reveals peaks corresponding to
the \jpsi, \psip and $\Upsilon$(1S,2S,3S) quarkonia states. The continuum
spectrum is well described by correlated heavy flavour ($D\bar{D}
\rightarrow \ee$ and $B\bar{B} \rightarrow \ee$) and Drell-Yan
contributions estimated using a PYTHIA simulation \cite{PYTHIA}, including
the detector acceptance and efficiencies.

The \psip differential cross section shown in Figure \ref{jpsi_psip}
is the first $p_T$-dependent result for an excited charmonium state at
RHIC. The \psip to \jpsi ratio is also shown and compared to results
from other experimental facilities 
\cite{HERA-B,CDF}. One can note the stability of this ratio for a vast 
collision energy and \pt. The up-to-date feed-down contributions to
\jpsi from \psip is $8.6 \pm 2.5\%$ and from $\chi_c$ is $<42\%$ (90\%
CL). These numbers are in agreement with world average results
\cite{Faccioli}.

\begin{figure}[ht]
\centering
\includegraphics[width=0.49\textwidth]{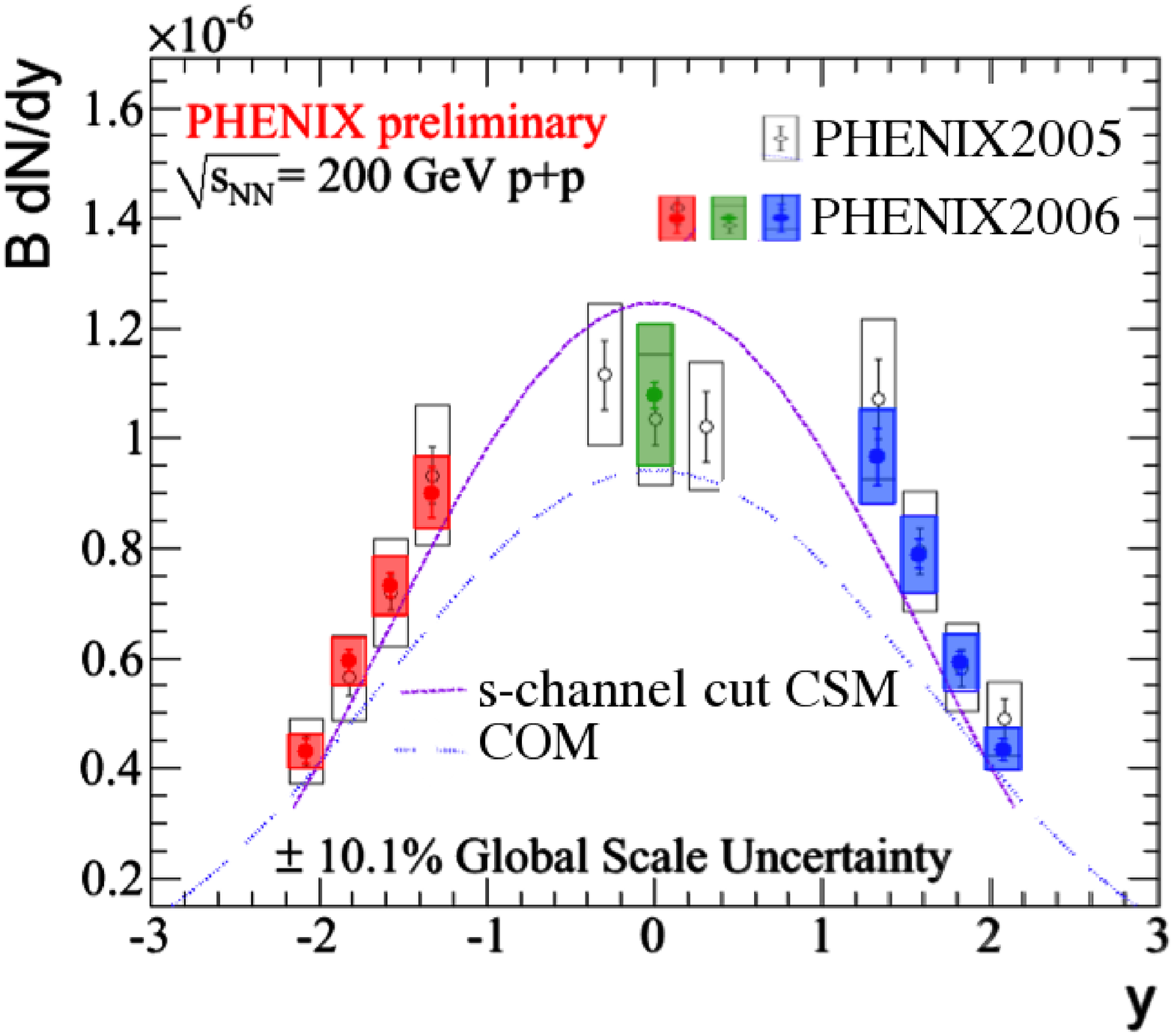}
\includegraphics[width=0.49\textwidth]{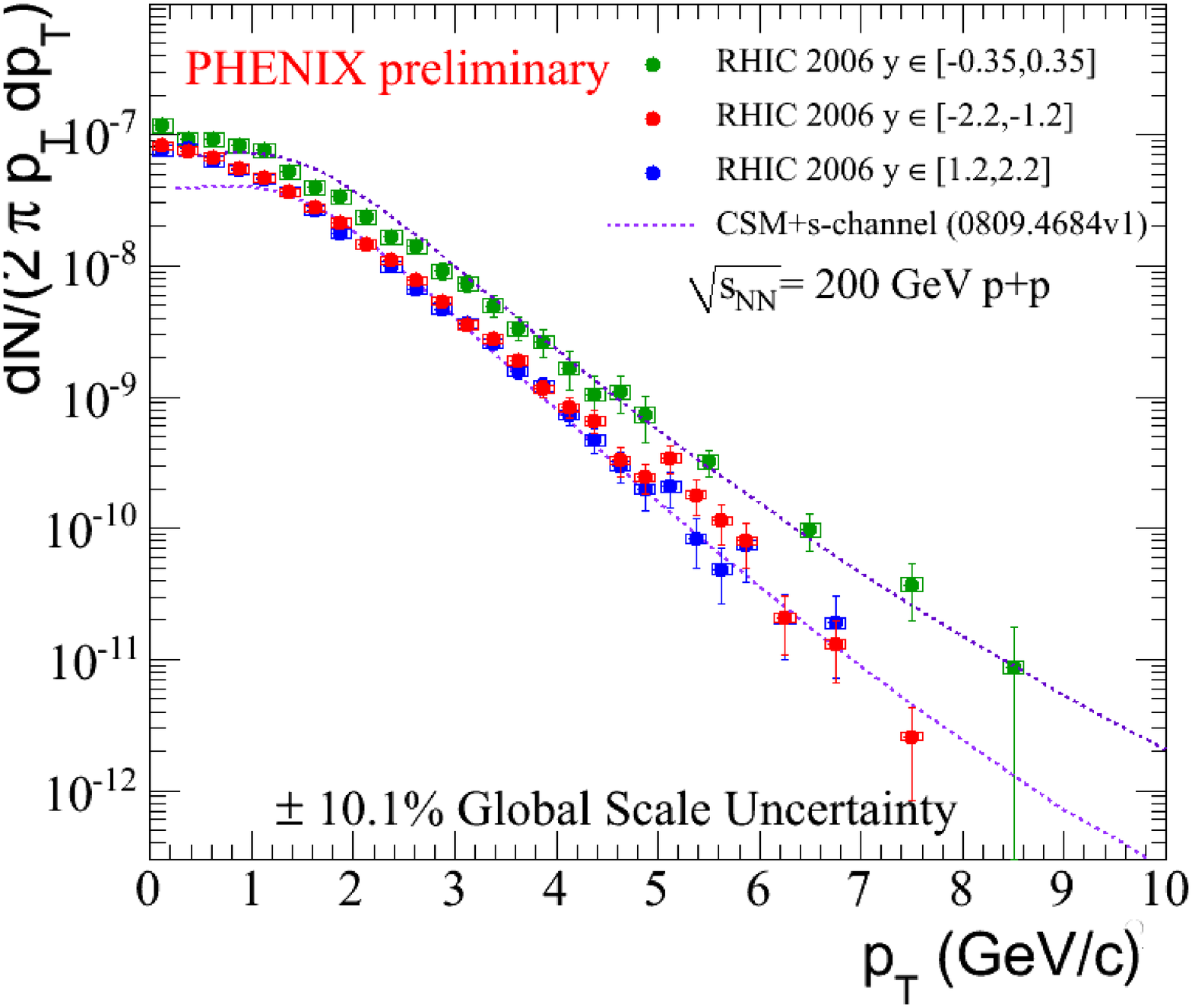}
\caption{Rapidity and \pt dependence of \jpsi production and comparison with
  non-relativistic QCD (NRQCD) Color Octet Model (COM) calculation\cite{Cooper},
  and an s-channel cut CSM calculation\cite{Lansberg}.}
\label{jpsi_yield}
\end{figure}

The new \jpsi rapidity dependence result (Figure \ref{jpsi_yield})
is in agreement with that already published \cite{PPG069}. The 
uncertainties are dominated by systematics. Both mid and forward
rapidity yields are well described by the s-channel cut Color Singlet
Model (CSM) \cite{Lansberg} using two parameters
fitted to CDF data at $\sqrt{s}=$1.8 TeV \cite{CDF} over $p_T < 10$ \gevc.

\begin{figure}[ht]
  \begin{minipage}{0.5\textwidth}
    \includegraphics[width=1.0\textwidth]{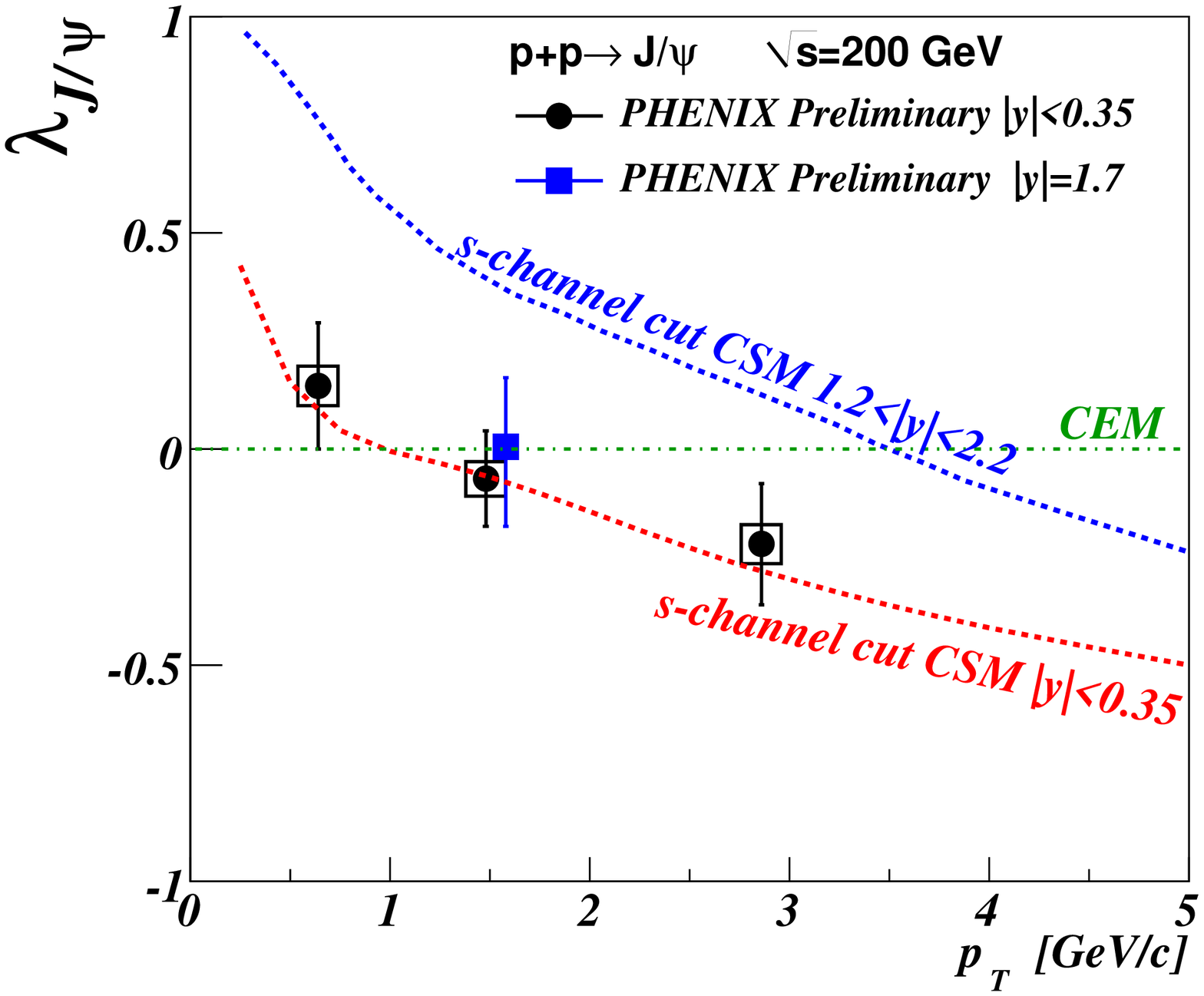}
    \caption{\jpsi polarization in the Helicity frame {\it versus}
      transverse momentum for mid and forward rapidity ranges, as well
      as predictions according to the s-channel cut
      CSM\cite{Lansberg_talk} and CEM.}
    \label{jpsi_pol}
  \end{minipage}
  \begin{minipage}{0.5\textwidth}
    \centering
    \includegraphics[width=0.9\textwidth]{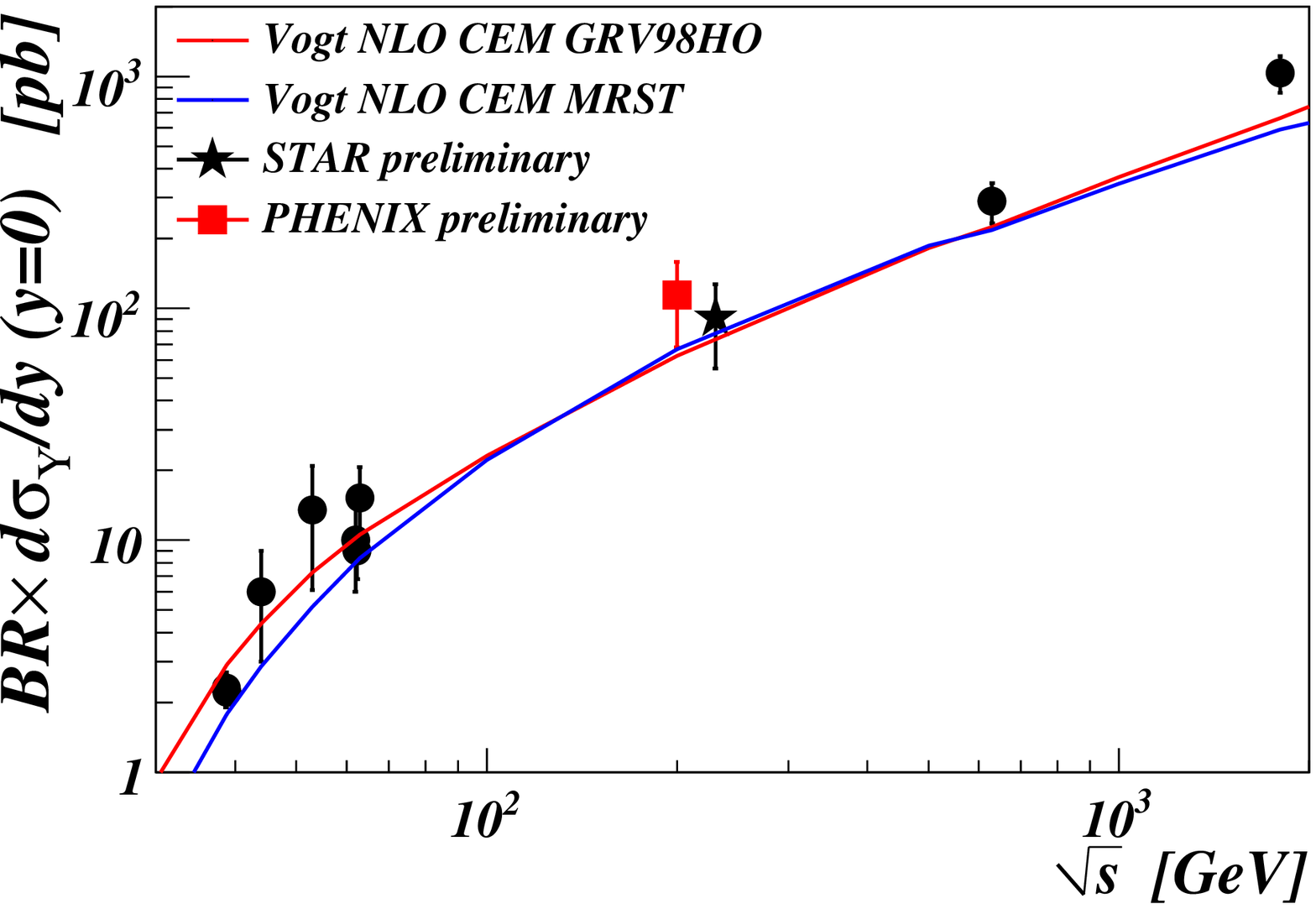}
    \includegraphics[width=0.9\textwidth]{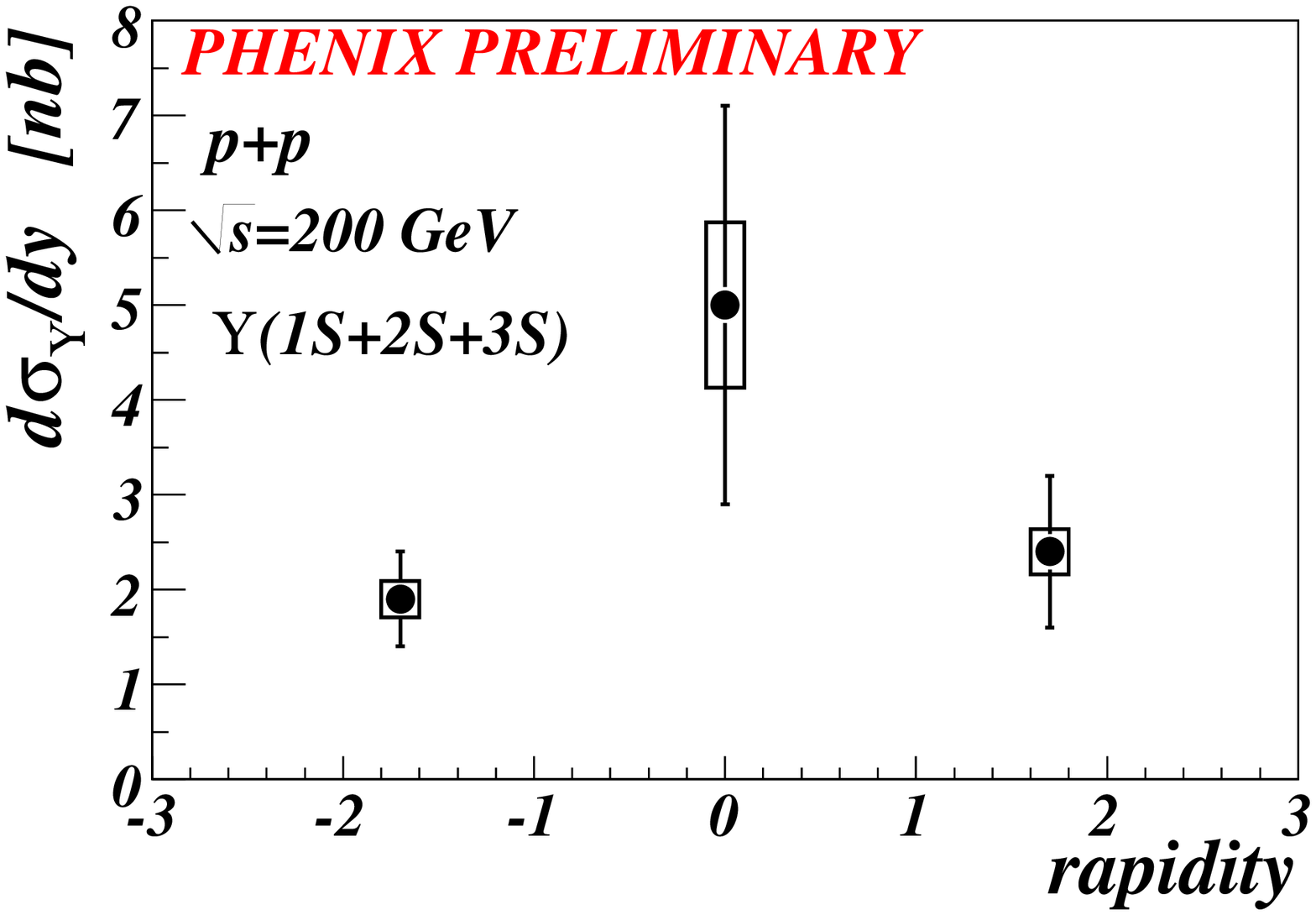}
    \caption{Collision energy and rapidity dependence of $\Upsilon$ yield.}
    \label{upsilon_yield}
  \end{minipage}
\end{figure}

PHENIX started a series of studies of the \pt dependence of
the decay angular distribution of the \jpsi. The polarization parameter
$\lambda$ is defined by \mbox{$\frac{dN}{d \cos \theta^{\prime}} = A
  \left( 1 + \lambda \cos\theta^{\prime} \right)$}, where
$\theta^{\prime}$ is the angle between the positive lepton and the
\jpsi momentum direction (Helicity frame). Initial inclusive \jpsi (direct + feed-down)
polarization {\it versus} \pt is shown in Figure
\ref{jpsi_pol}. Although the s-channel cut CSM only considers  
direct \jpsi (no feed-down) \cite{Lansberg_talk}, it is in good
agreement with the mid-rapidity results. However, the $p_T<5$ \gevc
result obtained at forward rapidity is $\sim 2 \sigma$ away from the
model prediction at that rapidity. Upcoming \pt dependent $\lambda$
measurements for the forward rapidity will be important to clarify
whether the s-channel cut CSM describes the data at forward rapidity
or not. Given the current uncertainties,
one cannot rule out the possibility of zero polarization, as expected by
the Color Evaporation Model (CEM) \cite{CEM}.

We counted 12 unlike-sign and one like-sign dielectron
pairs in the $\Upsilon$(1S+2S+3S) mass range $\in[8.5,11.5]~
GeV/c^2$ at mid-rapidity. A preliminary study of possible physical contributions in
this region indicates correlated open bottom and Drell-Yan can
contribute up to 15\% of the number of net counts and is accounted for
in the systematic errors. The corresponding $\Upsilon$ family cross 
section in the dielectron channel is \mbox{$\left.B
  \frac{d\sigma_{\Upsilon}}{dy}\right|_{|y|<0.35} =
114^{+46}_{-45}p$b.} This result follows the world trend and estimates
based on the CEM (Figure \ref{upsilon_yield})
\cite{Ramona_upsilon}. The rapidity dependence of the $\Upsilon$ cross
section, adding the preliminary results released at QM06
\cite{upsilon_qm06}, can be used to test production models for
bottomonium and estimate the total cross section. Given the smaller
relativistic contributions, NRQCD calculations for $\Upsilon$
at RHIC are supposed to be more reliable than for charmonium. The new
mid-rapidity measurement was also used as a baseline for the first
measurement of the $\Upsilon$ nuclear modification factor in Au+Au
collisions \cite{Ermias}.

\section{Cold Nuclear Matter (CNM) in $d$+Au collisions.}
\label{sec:dAu}

The fits performed to the published nuclear modification factors in $d$+Au
collisions ($R_{dA}$)  have been revisited taking into
account all systematic uncertainties \cite{dau_run3}. The new fits return a larger
uncertainty in the breakup cross sections. Cold nuclear matter
``extrinsic'' calculations, where one treats the production as a
2$\rightarrow$2 process ($g+g\rightarrow \jpsi+g$), have shown considerable 
differences in the breakup cross section compared to those obtained
using the standard procedure, using ``intrinsic'' kinematics with
2$\rightarrow$1 processes \cite{Andry}.

The integrated luminosity accumulated in $d$+Au collisions during the
2008 Run was about thirty times larger than that used in \cite{dau_run3}. The
first nuclear modification factor measurements from the 2008 Run were
performed using as a reference the yield in the 60-80\%
centrality range,

\begin{eqnarray}
  \label{eq:rcp}
  R_{cp} = \frac{\left(\frac{dN_{J/\psi}}{dy}/N_{coll}
    \right)}{\left(\frac{dN^{60-80\%}_{J/\psi}}{dy}/N^{60-80\%}_{coll}
    \right)}.
\end{eqnarray}

\begin{figure}[ht]
\centering
\includegraphics[width=0.4\textwidth]{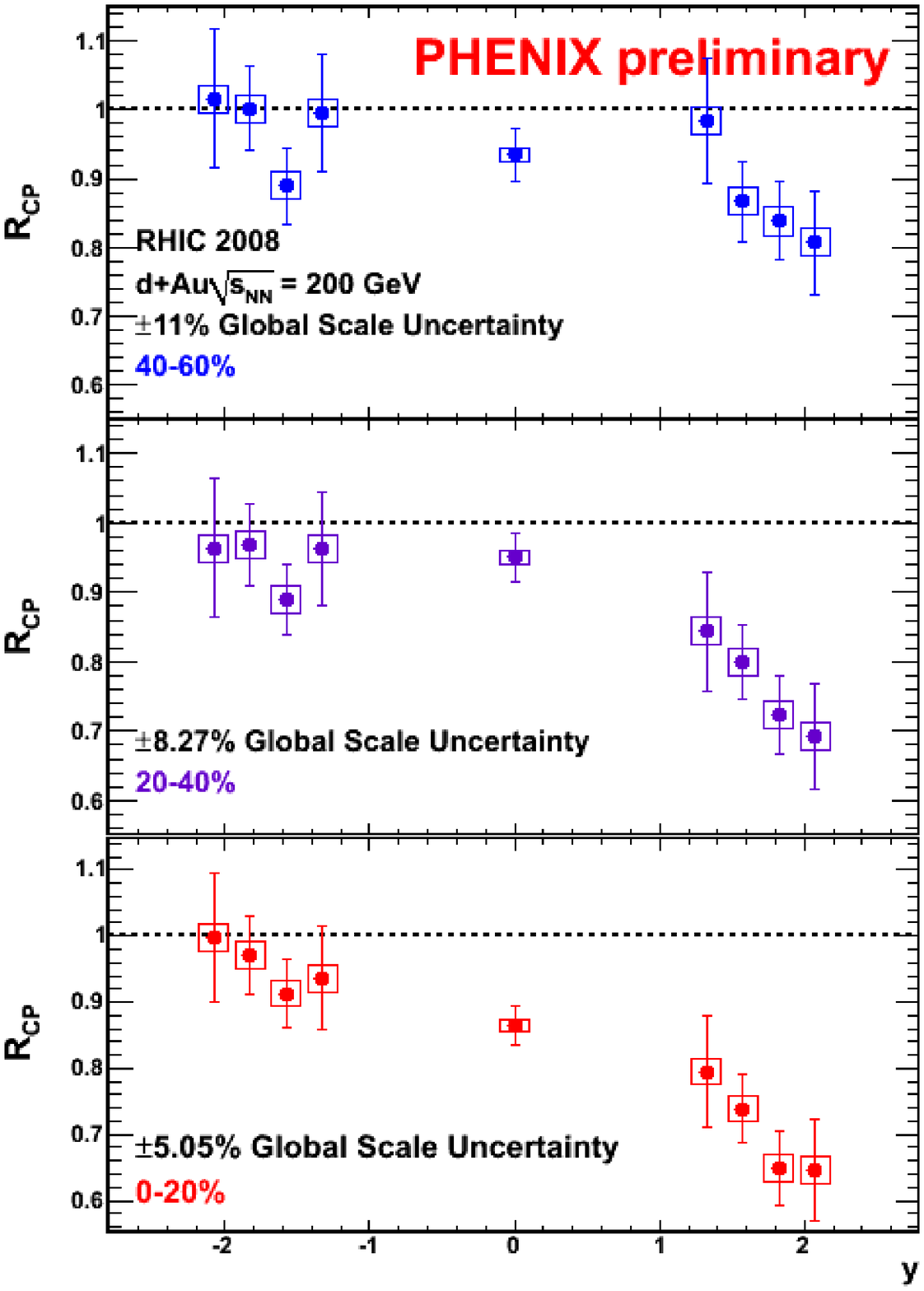}
\includegraphics[width=0.4\textwidth]{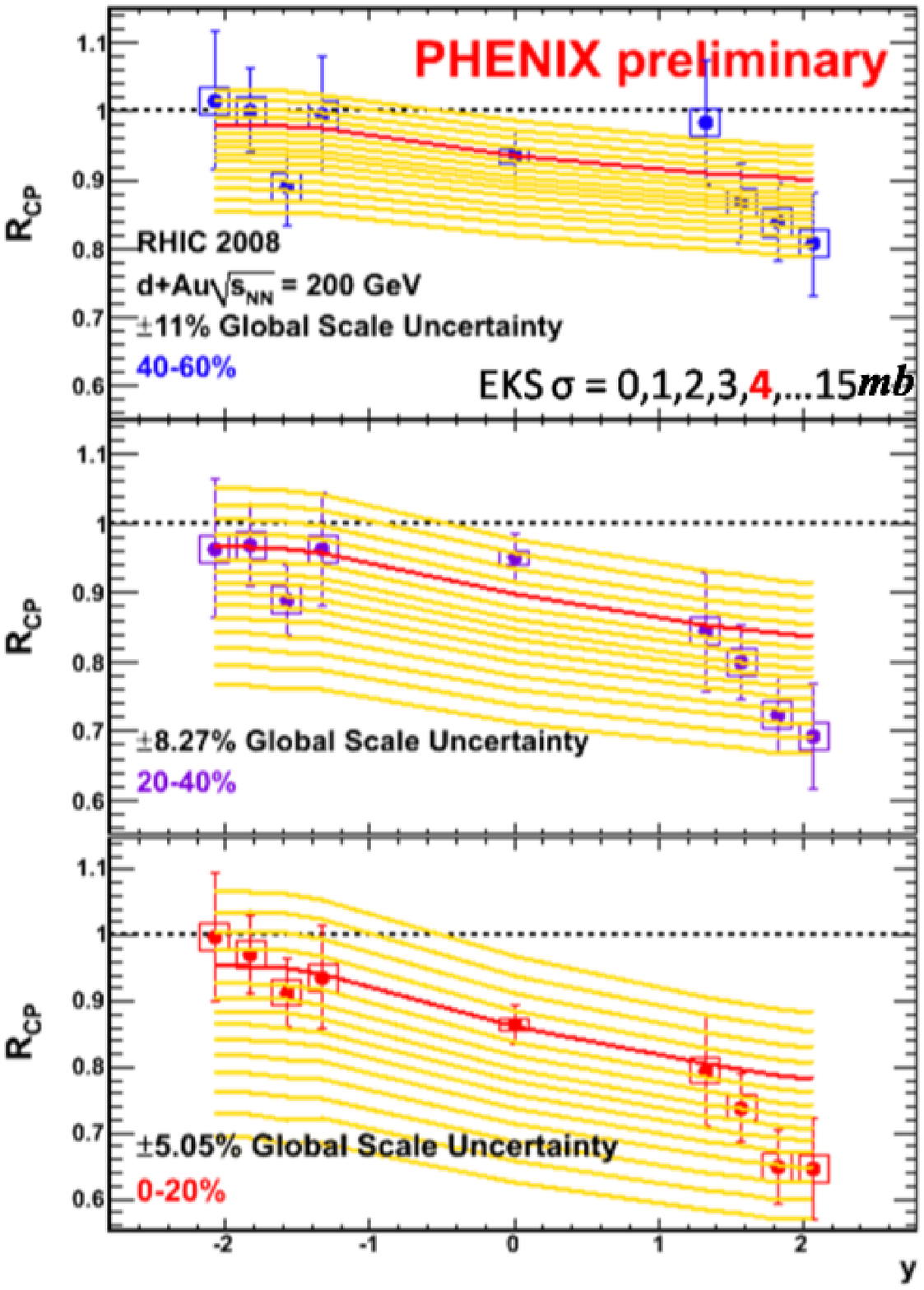}
\caption{Rapidity dependence of the modification factor as defined in
  (\ref{eq:rcp}) and comparison with that from EKS98 parton
  modification distributions for different breakup cross
  sections\cite{Ramona}.}
\label{rcp}
\end{figure}

Many systematic uncertainties cancel out in a $R_{cp}$
measurements. Figure \ref{rcp} shows the rapidity dependence of
$R_{cp}$ for three centrality ranges. The negative rapidity region, which
represents large $x$ (in the anti-shadowing region) of the gluon
distribution, shows no nuclear effects within uncertainties, but a
considerable suppression is observed at central 
rapidity and at forward rapidity. One observes that the strong
suppression at forward rapidity and the lack of suppression at
backward rapidity cannot be described by the existing 
nuclear modified gluon distributions (EKS98 shown in Figure \ref{rcp}, EPS08
and NDSG) for any fixed breakup cross section \cite{Ramona}. This
observation suggests that additional CNM effects besides 
the standard shadowing and hadronic interactions must be important.

Upcoming $R_{dA}$ results, using measurements  in \pp collisions as a
reference, will extend this study of the CNM effects.

%% end of main text

%\section*{Acknowledgments} % please insert, comment out or delete if not needed
%This is where one places acknowledgments for funding bodies etc., if needed.
%For the large collaborations, this is listed once and for all, together with 
%the author lists etc. in the proceedings back-material.

 % do not change 
\end{document}